\pdfoutput=1
\documentclass[aps,prl,twocolumn,groupedaddress,showpacs]{revtex4}

\usepackage{graphicx}
\usepackage{amsmath}

\begin{document}

\title{The Electronic Specific Heat of Ba$_{1-x}$K$_x$Fe$_2$As$_2$ from 2K to 380K.}

\author{J.G. Storey$^1$, J.W. Loram$^1$, J.R. Cooper$^1$, Z. Bukowski$^2$ and J. Karpinski$^2$}

\affiliation{$^1$Cavendish Laboratory, University of Cambridge, Cambridge CB3 0HE, U.K.}

\affiliation{$^2$Laboratory for Solid State Physics, ETH Zurich, Zurich, Switzerland}

\date{\today}

\begin{abstract}
Using a differential technique, we have measured the specific heats of polycrystalline Ba$_{1-x}$K$_x$Fe$_2$As$_2$ samples with $x$ = 0, 0.1 and 0.3, between 2K and 380K and in magnetic fields 0 to 13 Tesla. From this data we have determined the electronic specific heat coefficient $\gamma$($\equiv C_{el}/T$) over the entire range for the three samples. The most heavily doped sample ($x$ = 0.3) exhibits a large superconducting anomaly $\Delta\gamma(T_c)$ $\sim$ 48 mJ/mol K$^2$ at $T_c$ = 35K, and we determine the energy gap, condensation energy, superfluid density and coherence length. In the normal state for the $x$ = 0.3 sample, $\gamma$ $\sim$ 47 mJ/mol K$^2$ is constant from $T_c$ to 380K. In the parent compound ($x$ = 0) there is a large almost first order anomaly at the spin density wave (SDW) transition at $T_o$ = 136K. This anomaly is smaller and broader for  $x$ = 0.1. At low $T$, $\gamma$ is strongly reduced by the SDW gap for both $x$ = 0 and 0.1, but above $T_o$, $\gamma$ for all three samples are similar.
\end{abstract}


\maketitle

The electronic specific heat contains a wealth of quantitative information on the electronic spectrum over an energy region $\pm$100meV about the Fermi level, crucial to the understanding of high temperature superconductivity. Measurements of the electronic specific heat played an important role in revealing key properties of the copper-oxide based `cuprate' high-temperature superconductors (HTSCs). Some examples include the normal-state ``pseudogap''\cite{OURWORK2,LORAMPG,LORAMU0,ENTROPYDATA2}, the bulk sample inhomogeneity length scale\cite{SCFLUC}, and more recently the degree to which the superconducting transition temperature is suppressed due to superconducting fluctuations\cite{TALLONFLUCS}. 
In this work we extend such measurements to the recently discovered iron-arsenide based `pnictide' HTSCs. Here we present results obtained from polycrystalline samples of Ba$_{1-x}$K$_x$Fe$_2$As$_2$ ($x$ = 0, 0.1 and 0.3) using a high-resolution differential technique\cite{LORAMCALORIMETER}.

With this technique we directly measure the difference in the specific heats of a doped sample and the undoped reference sample (BaFe$_2$As$_2$). This eliminates most of the large phonon term from the raw data and yields a curve which is dominated by the difference in electronic terms between the sample and reference. After making a small correction for any residual phonon term, this difference in electronic specific heats can be determined with a resolution of $\sim$ 0.1 mJ/mol K$^2$ at temperatures from 1.8K to 380K and magnetic fields from 0 to 13T. During a measurement run the total specific heats of the sample and reference are also measured. 

The polycrystalline samples of Ba$_{1-x}$K$_x$Fe$_2$As$_2$ were prepared by a solid state reaction method similar to that reported by Chen \textit{et al}\cite{CHEN}. First, Fe$_2$As, BaAs, and KAs were prepared from high purity As (99.999\%), Fe (99.9\%), Ba (99.9\%) and K (99.95\%) in evacuated quartz ampoules at 800, 650 and 500$^\circ$C respectively. Next, the terminal compounds BaFe$_2$As$_2$ and KFe$_2$As$_2$ were synthesized at 950 and 700$^\circ$C respectively, from stoichiometric amounts of BaAs or KAs and Fe$_2$As in alumina crucibles sealed in evacuated quartz ampoules. Finally, samples of Ba$_{1-x}$K$_{x}$Fe$_2$As$_2$ with $x$ = 0.1 and 0.3 were prepared from appropriate amounts of single-phase BaFe$_2$As$_2$ and KFe$_2$As$_2$. The components were mixed, pressed into pellets, placed into alumina crucibles and sealed in evacuated quartz tubes. The samples were annealed for 50 h at 700$^\circ$C with one intermittent grinding, and were characterized by room temperature powder X-ray diffraction using Cu K$_\alpha$ radiation. The diffraction patterns were indexed on the basis of the tetragonal ThCr$_2$Si$_2$ type structure (space group I4/mmm). Lattice parameters calculated by a least-squares method agree well with those reported by Chen \textit{et al}\cite{CHEN}. Traces of FeAs as an impurity were detected for compositions with $x$ = 0.1 and 0.3. The samples for heat capacity measurement weighed $\sim$ 0.8g.

\begin{figure}
\centering
\includegraphics[width=\linewidth,clip=true,trim=0 0 0 0]{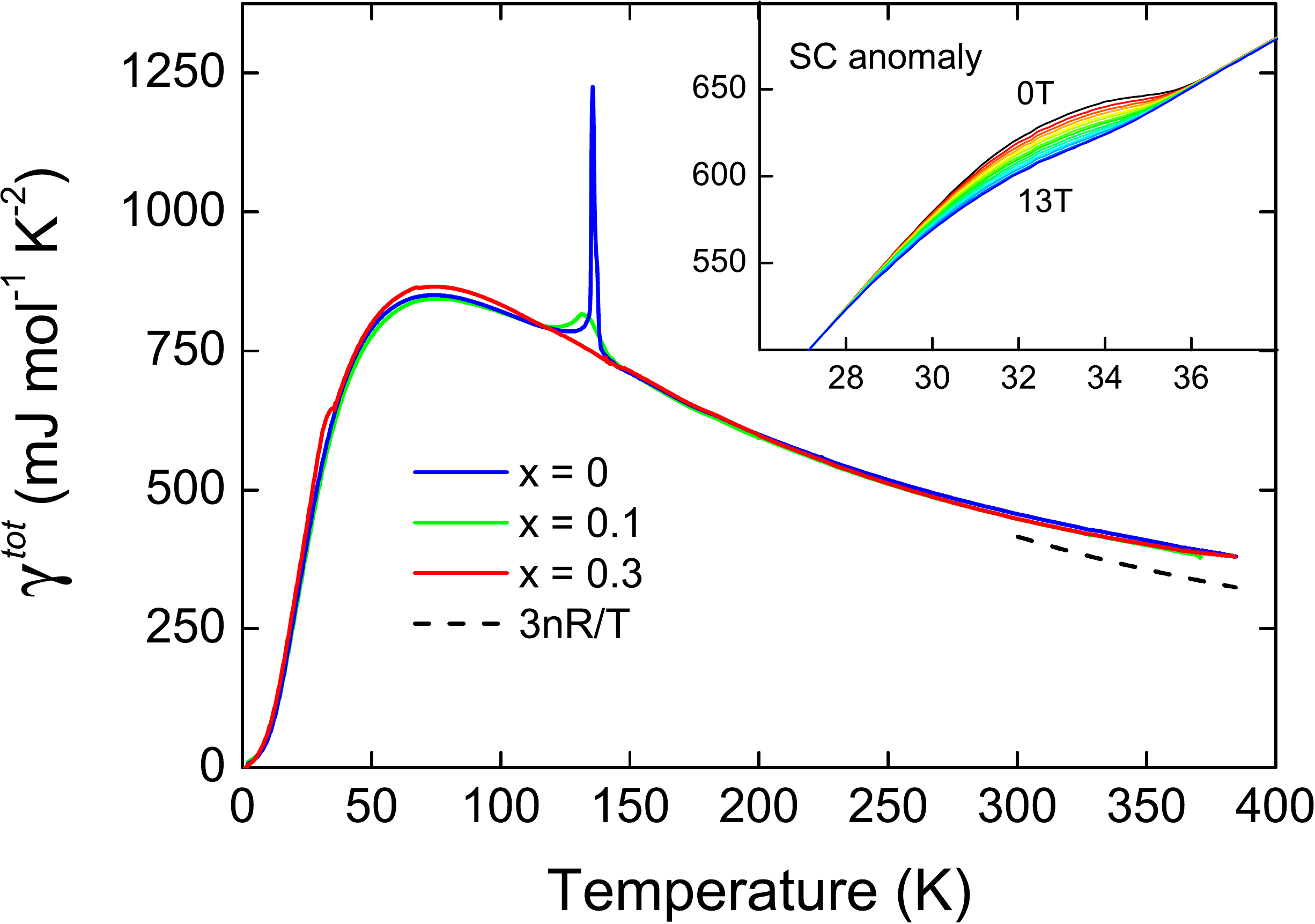}
\caption{(Color online) $\gamma^{tot}\equiv C^{tot}/T$  of Ba$_{1-x}$K$_x$Fe$_2$As$_2$ for $x$ = 0, 0.1 and 0.3. The dotted line shows the high temperature limiting value of the lattice contribution, $3nR/T$. The inset shows the suppression of the superconducting anomaly in magnetic fields from 1 to 13T in 1T increments.}
\label{TOTALCFIG}
\end{figure}

\begin{figure}
\centering
\includegraphics[width=70mm,clip=true,trim=0 0 0 0]{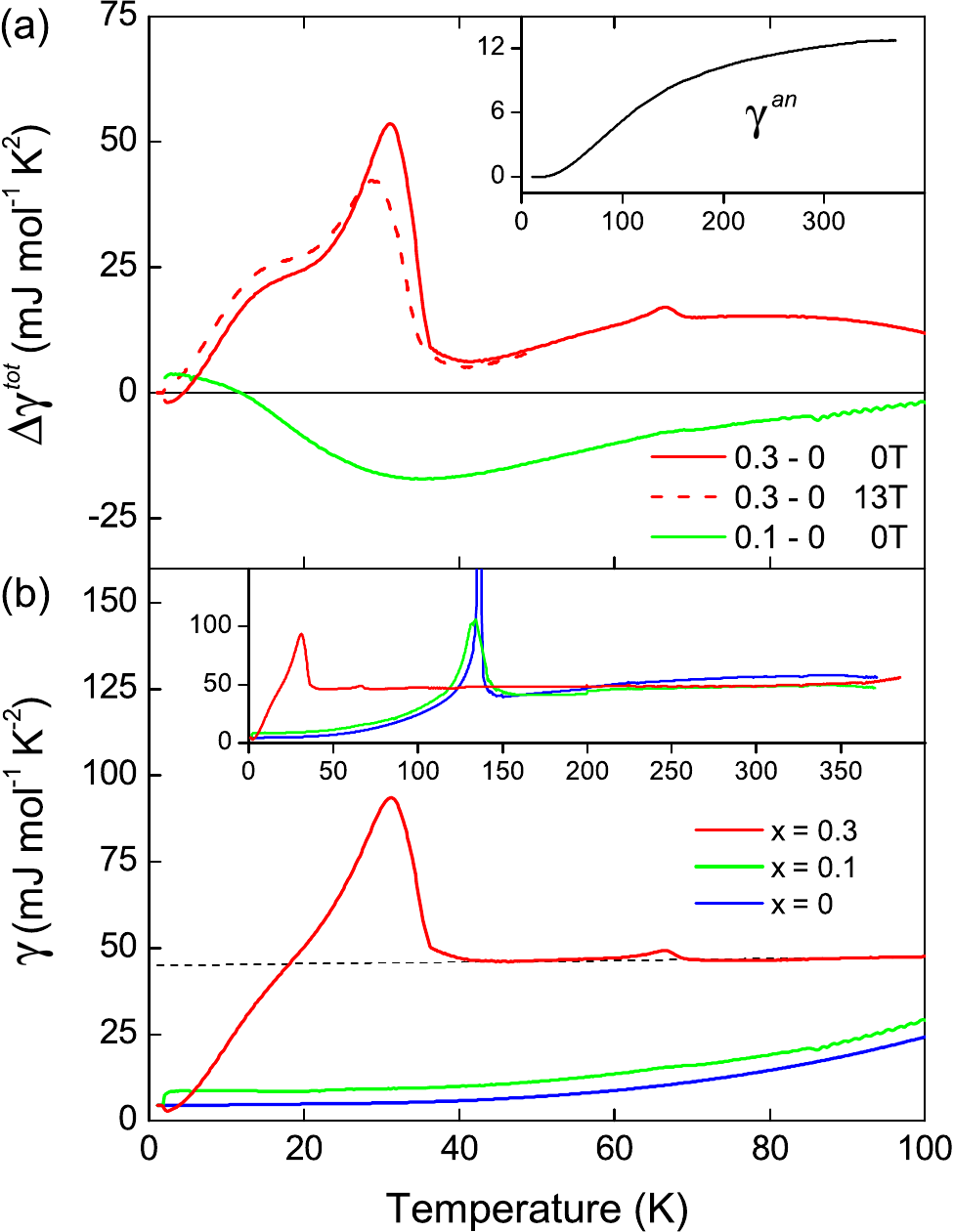}
\caption{(Color online) (a) The directly measured raw difference in specific heats between the doped samples and undoped reference in 0 and 13T fields. The inset shows the anharmonic phonon term $\gamma^{an}$. (b) The electronic specific heat of Ba$_{1-x}$K$_{x}$Fe$_2$As$_2$ for $x$ = 0, 0.1 and 0.3.}
\label{GAMMAFIG}
\end{figure}

The total specific heats of the three samples are plotted as $\gamma^{tot}\equiv C^{tot}/T$ in Fig.~\ref{TOTALCFIG}. In the $x$ = 0 sample there is a sharp and almost first order anomaly at the magneto-structural transition at $T_0$ = 136K, in agreement with published single crystal data\cite{DONG}. For the $x$ = 0.1 sample the corresponding anomaly at $T_0$ = 135K is broader and considerably reduced in magnitude, in agreement with the data of Rotter \textit{et al}.\cite{ROTTER} who have tracked $T_0$ in the specific heat down to 105K in a sample with $x$ = 0.2. The magnetic field dependence of this anomaly is extremely weak for both samples. We see no evidence for a structural transition in the $x$ = 0.3 sample (the very weak anomaly at 67K in the differential data (Fig.~\ref{GAMMAFIG}(a)) is probably due to an FeAs impurity phase\cite{SELTE}). The superconducting transition in this sample is clearly visible in $\gamma^{tot}$ at 35K, and the field dependence is shown in the inset to Fig.~\ref{TOTALCFIG}.

In the temperature range 2 $-$ 8K, $\gamma^{tot}=\gamma(0)+\beta(0)T^2$ at zero field with $\gamma(0)$ = 5.1, 9.3 and 1.8 mJ/mol K$^2$ and $\beta(0)$ = 0.45, 0.37 and 0.57 mJ/mol K$^4$ for the $x$ = 0, 0.1 and 0.3 samples respectively. On the assumption that $\beta(0)$ is entirely due to phonons we obtain Debye temperatures  $\theta_D(0)$ = 289, 298 and 257K for the three samples. We cannot, however, ignore the possibility that for the superconducting 0.3 sample $\beta(0)$ is enhanced by an electronic component $\propto T^2$, leading to too low a value of $\theta_D(0)$. This would explain the reduced value of  $\theta_D(0)$ for this sample which contradicts the trend seen in $\Delta\gamma^{tot}$ (Fig.~\ref{GAMMAFIG}(a)) that acoustic phonon frequencies, which dominate the phonon specific heat in this temperature region, increase with doping. If the 0 and 0.1 samples have an initial electronic $T^2$ term in $\gamma(T)$ we would expect this to have a magnitude $\sim\gamma_n T^2 /T_o^2 \sim$ 0.003 mJ/mol K$^2$ if it is controlled by the same energy scale as $T_0$. This contribution to $\beta$ would be negligible compared with the measured $\beta(0)$.

We next discuss the separation of electronic and phonon contributions from the total specific heat over a wider temperature range. The total specific heat coefficient is given by $\gamma^{tot}=\gamma+\gamma^{ph}+\gamma^{an}$
where the electronic term $\gamma=C_{el}/T$, the harmonic phonon term $\gamma^{ph}=C_v/T$ and the anharmonic phonon (dilation) term $\gamma^{an}=(C_p-C_v)/T$\cite{ASHCROFT}. $C_p$ and $C_v$ are the lattice heat capacities at constant pressure and volume respectively. The anharmonic term is given by $\gamma^{an}=VB\beta^2$\cite{ASHCROFT} where $V$ is the molar volume, $B$ is the bulk modulus and $\beta$ is the volume expansion coefficient. We assume that $\gamma^{an}$ is doping independent and use $V$ = 61 cm$^3$/mol\cite{ROTTER}, $B \sim$ 0.80$\times$10$^8$ mJ/cm$^3$\cite{KIMBER}, and $\beta$(300K) $\sim$ 50$\times$10$^{-6}$ /K\cite{BUDKO} to obtain a room temperature value $\gamma^{an}$(300K) $\sim$ 12 mJ/mol K$^2$ for each sample.  To a very good approximation $\beta(T)\propto C_v(T)$\cite{ASHCROFT} and thus
$\gamma^{an}(T) = [C_v(T)/C_v(300K)]^2\cdot\gamma^{an}(300K)$.
A plot of $\gamma^{an}(T)$ is shown in the inset to Fig.~\ref{GAMMAFIG}(a).

After correcting for $\gamma^{an}(T)$, the electronic term $\gamma$ can be determined at high temperatures where $C_v$ is close to its saturation value $3nR$ (where $n$ is the number of atoms/formula unit) shown by the dotted line in Fig.~\ref{TOTALCFIG}. To extend the useful range of the high temperature region we exploit the fact that the Debye temperature $\theta_D(T)$ deduced from $C_v$ is generally only weakly $T$-dependent for $T>0.5\theta_D(\infty)$ (or $\sim$ 180K for the present materials). With this constraint we obtain limiting high temperature Debye temperatures $\theta_D(\infty)$ = 368, 360 and 356K and high-$T$ values $\gamma(300K)$ = 55, 47 and 47 mJ/mol K$^2$ for the 0, 0.1 and 0.3 samples.

Differential measurements (see Fig.~\ref{GAMMAFIG}(a)) between each sample and the $x$ = 0 reference give $\Delta\gamma^{tot}=\Delta\gamma+\Delta\gamma^{ph}$ (assuming $\Delta\gamma^{an}$ = 0). It is clear from Figs.~\ref{TOTALCFIG} and \ref{GAMMAFIG}(a) that the phonon terms for the three samples are very similar. The broad negative peak at 35K seen in $\Delta\gamma^{tot}(0.1,0)=\gamma^{tot}(x=0.1)-\gamma^{tot}(x=0)$, and which can also be inferred in the data for $\Delta\gamma^{tot}$(0.3,0) in the same temperature region, is compatible with fractional increases of 0.018 and 0.030 in the acoustic phonon frequencies for the 0.1 and 0.3 samples relative to the $x$ = 0 reference.  Since we expect the difference of electronic terms $\Delta\gamma$(0.1,0) for the two non-superconducting samples to be only weakly $T$-dependent (at least below 100K), we obtain   $\Delta\gamma^{ph}(0.1,0)=\Delta\gamma^{tot}(0.1,0)-\Delta\gamma(0.1,0)$ using the low temperature value $\Delta\gamma(0.1,0) = -4.2$ mJ/mol K$^2$.  To ensure that the resulting   $\Delta\gamma^{ph}$(0.1,0) has a $T$-dependence compatible with that of a phonon spectrum it was modelled with a histogram for the difference phonon spectrum, as discussed previously\cite{LORAM93}. Making this phonon correction removes the broad negative peak in $\Delta\gamma^{tot}$ at 35K for this sample (Fig.~\ref{GAMMAFIG}(b)). As shown in the same figure, subtracting a phonon term $\Delta\gamma^{ph}$(0.3,0) = 1.68$\Delta\gamma^{ph}$(0.1,0) also removes the negative peak in $\Delta\gamma^{tot}$(0.3,0). Following this correction, the electronic  $\Delta\gamma$(0.3,0) has an additional negative $T$-dependence given by $\sim$ $-20\times10^{-6}T^3$  mJ/mol K$^2$ in the range 40 - 110K. We conclude below that $\gamma_n(T)$ for the 0.3 sample is $T$-independent, and attribute this negative $T$-dependence to a corresponding positive term in the electronic specific heat coefficient $\gamma(x=0)$ of the undoped	 sample in the SDW phase.

After correcting for $\Delta\gamma^{ph}$(0.3,0) we find that just above $T_c$, $\gamma(x=0.3) \sim$ 47 mJ/mol K$^2$. This is very close to the value $S/T(T_c)$, where $S = \int_0^T{\gamma(T)dT}$ is the electronic entropy. By definition, $S/T(T^\prime)$ is the average value of $\gamma$ below $T^\prime$. From conservation of entropy  $S/T(T_c)$ is also the average value of the underlying normal state $\gamma_n(T)$ below $T_c$. Therefore the result above is consistent with a $T$-independent $\gamma_n(T)$ for $T<T_c$. In addition, this value for $\gamma_n(T)$ below $T_c$ is close to its high temperature value estimated above, and it is reasonable to assume that for this sample $\gamma_n(T)$ is $T$-independent over the entire temperature range. Subtracting this constant term from $\gamma^{tot}(x=0.3)$ gives $\gamma^{ph}(x=0.3)$, which we then fit with a nine-term histogram for the phonon spectrum\cite{LORAM93}. Finally, subtracting $\Delta\gamma^{ph}$(0.3,0) and $\Delta\gamma^{ph}$(0.3,0.1) from $\gamma^{ph}(x=0.3)$ gives $\gamma^{ph}(x=0)$ and $\gamma^{ph}(x=0.1)$.

\begin{figure}
\centering
\includegraphics[width=70mm,clip=true,trim=0 0 0 0]{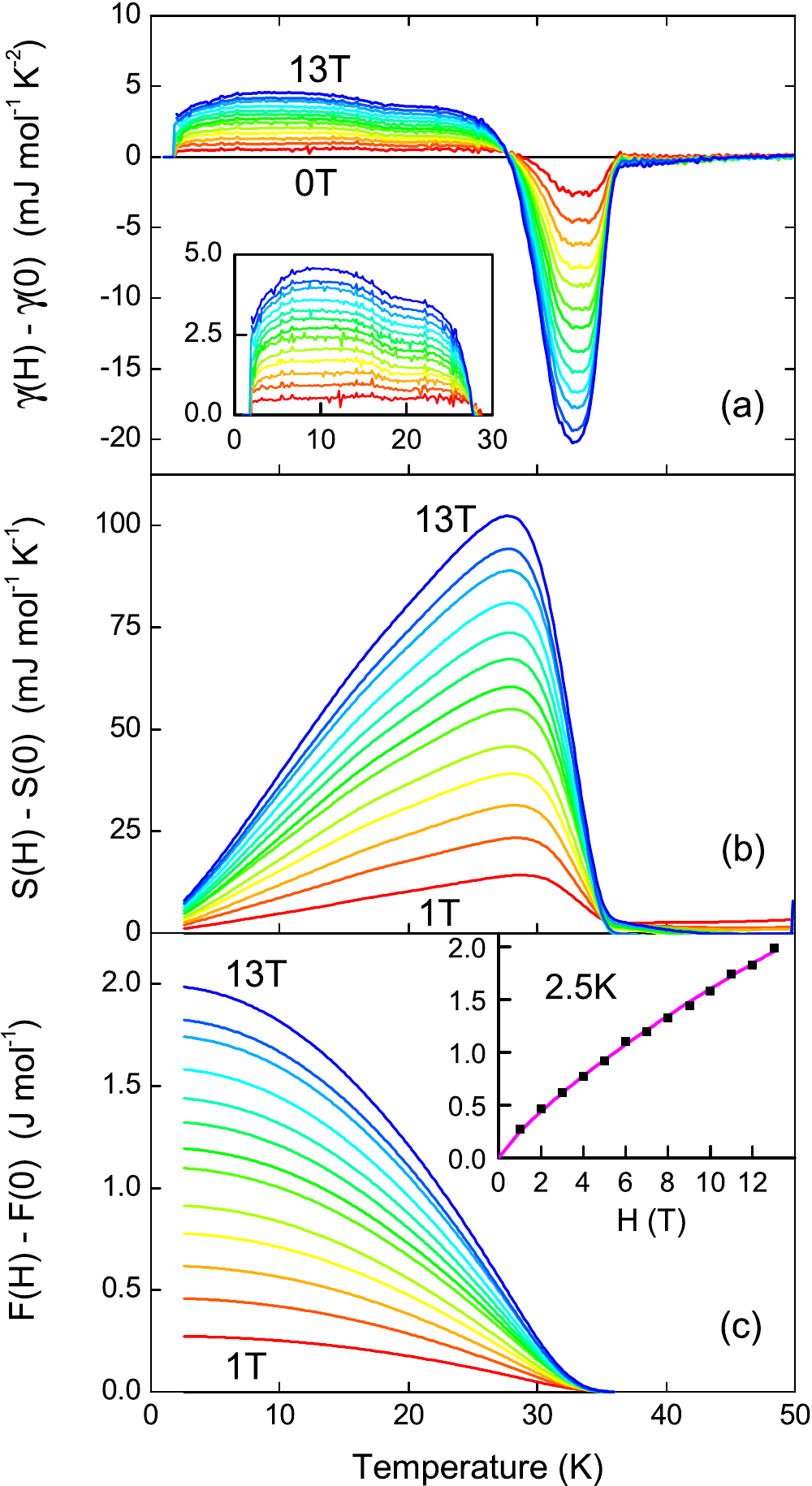}
\caption{(Color online) (a) Magnetic field dependent change in electronic specific heat $\gamma(H)-\gamma(0)$ vs temperature, in 1T increments, from the data in the inset of Fig.~\ref{TOTALCFIG}. (Inset) Enlargement of the 0 to 30K region. (b) Field dependence of the electronic entropy, $S(H)-S(0)$, calculated from (a) using Eqn.~\ref{ENTROPYEQN}. (c) Field dependence of the free energy, $F(H)-F(0)$, calculated from (b) using Eqn.~\ref{FEQN}. (Inset) A fit to $F(H)-F(0)$ at 2.5K using Eqn.~\ref{HAOCLEMEQN}.}
\label{GSFFIG}
\end{figure}

The electronic specific heats of the three samples are shown in Figure~\ref{GAMMAFIG}(b). Above 150K, $\gamma$ is at most only weakly $T$-dependent and almost independent of doping.  At low temperatures
$\gamma$ for the $x$ = 0 and 0.1 samples is heavily reduced by factors of  $\sim$ 11 and 5 respectively due to the opening of a spin density wave (SDW) gap below $T_0$ $\sim$ 136K, and increases as $\sim$ $20\times10^{-6}T^3$  mJ/mol K$^2$ in the range 40 $-$ 110K. Band splitting and signs of partial gapping of the Fermi surface have been observed in the SDW state of BaFe$_2$As$_2$ by angle-resolved photoemission spectroscopy\cite{YANG}.
$\gamma$ for the $x$ = 0.3 sample is dominated by the large superconducting anomaly with $\Delta\gamma(T_c)$ $\approx$ 48 mJ/mol K$^2$,  a normal-state $\gamma_N$(0) of about 47 mJ/mol K$^2$ and $\gamma(0)$ = 1.8 mJ/mol K$^2$  . Measurements on a more overdoped crystal with $x$ = 0.4 by Mu \textit{et al}.\cite{MU} show an even larger $\Delta\gamma(T_c)$ $\approx$ 100 mJ/mol K$^2$ and $\gamma_N$(0) $\approx$ 63 mJ/mol K$^2$, implying a further growth in the density of states with doping. For our $x$ =0.3 sample the superconducting condensation energy $U(0)$ = 10.6 J/mol, as determined from $\int_0^{T_c}\left(S_n-S_s\right)dT$, where $S_n$ and $S_s$ are the normal and superconducting state entropies respectively.
In Fig.~\ref{GAMMAFIG}(b), the curvature in $\gamma$ for $x$ =0.3 between 10 and 20K is reduced compared to the raw $\Delta\gamma^{tot}$ data (Fig.~\ref{GAMMAFIG}(a)) by the phonon correction described above. Evidence that the remaining curvature in $\gamma$ may reflect a genuine anomalous $T$-dependence of the electronic term comes from the observation of similar anomalous curvature between 15 and 20K in the field dependence  $\Delta\gamma(H)=\gamma(H,T)-\gamma(0,T)$ shown in the inset to Fig.~\ref{GSFFIG}(a). Since the phonon term is $H$-independent, this anomalous curvature in   $\Delta\gamma(H)$ can only be of electronic origin, and may signal the presence of a second energy gap as inferred by Mu \textit{et al}\cite{MU}. 
\begin{figure}
\centering
\includegraphics[width=70mm,clip=true,trim=0 0 0 0]{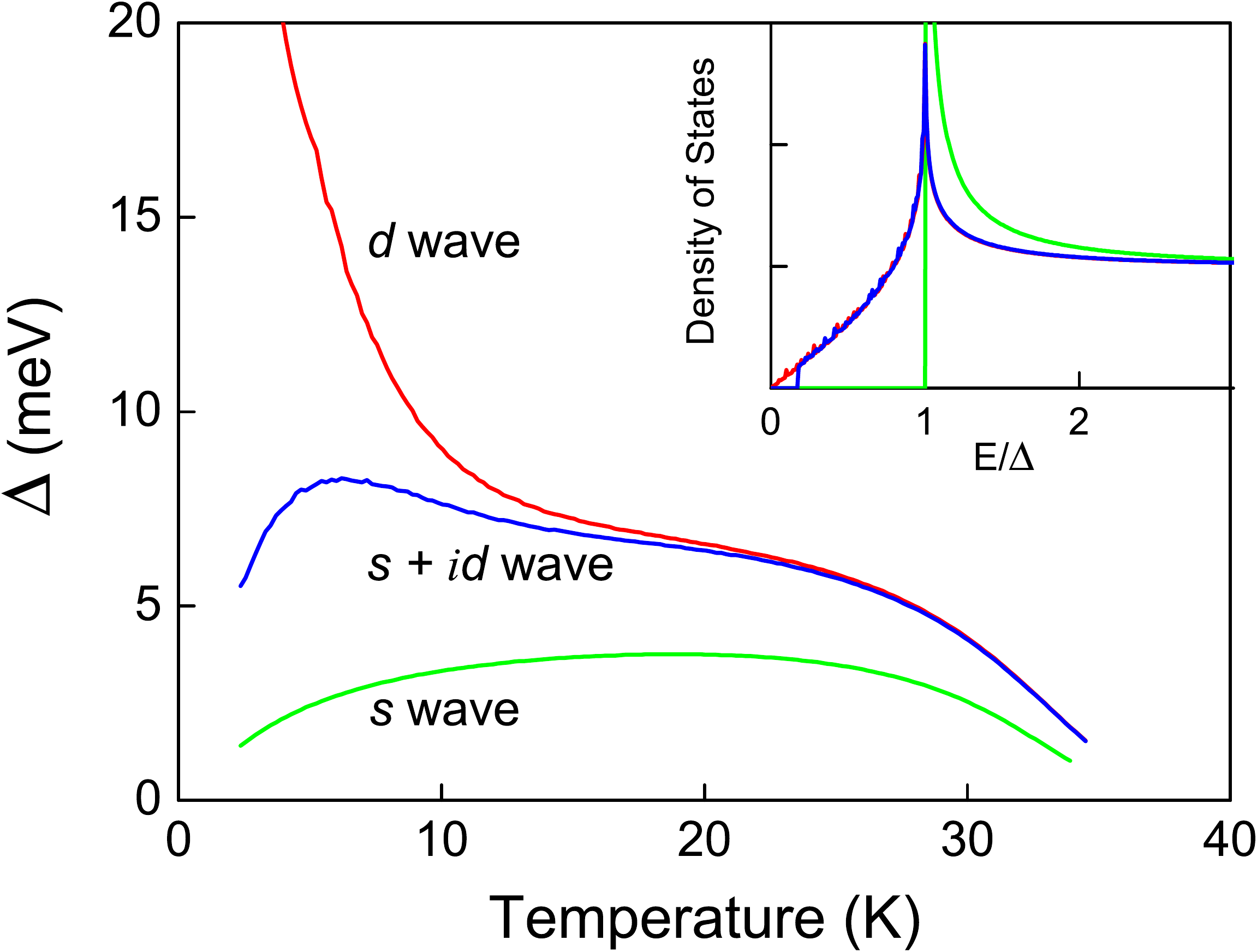}
\caption{(Color online) Temperature dependence of the superconducting gap extracted by comparing the experimentally determined $S/T$ with that calculated from models of the density of states (shown in the inset) given by Eqns.~\ref{SWAVEEQN}-\ref{SDWAVEEQN}.}
\label{GAPFIG}
\end{figure} 

To determine the magnitude and temperature dependence of the superconducting gap $\Delta$ for $x$ = 0.3  (shown in Fig.~\ref{GAPFIG}) we first subtract from $\gamma$ the small residual $\gamma(0)$ = 1.8 mJ/mol K$^2$ and then match the corrected  $S/T$ with that calculated assuming the following models for the density of states (see Fig.~\ref{GAPFIG} inset):

$s$-wave
\begin{equation}
N(E)=E/\sqrt{E^2-\Delta^2}
\label{SWAVEEQN}
\end{equation}

$d$-wave
\begin{equation}
N(E)=\frac{1}{N_\theta}\sum_\theta E/\sqrt{E^2-\Delta^2\cos^2 2\theta}
\label{DWAVEEQN}
\end{equation}

and $s+id$-wave
\begin{equation}
N(E)=\frac{1}{N_\theta}\sum_\theta E/\sqrt{E^2-0.98\Delta^2\cos^2 2\theta-0.02\Delta^2}
\label{SDWAVEEQN}
\end{equation}
where the summations over $\theta$ are from 0 to 45$^\circ$ and $N_\theta$ is the number of $\theta$ values.
Although such modelling is too crude to pin down the exact nature of the gap, the rapid increase in $\Delta$ below 15K in the case of the assumed $d$-wave gap function, shows that nodes on the Fermi surface are incompatible with the $T$-dependence of $\gamma$ observed at low temperature, even in the presence of weak pair breaking. The $s+id$-wave gap function produces a more reasonable $\Delta(T)$. This function completely gaps the Fermi surface but is strongly anisotropic. Such anisotropy could arise from the presence of multiple gaps on different sheets of the Fermi surface, or from an intrinsically anisotropic gap(s).

We turn now to the information contained in the magnetic field dependence of $\gamma$.
The inset to Fig.~\ref{TOTALCFIG} shows the suppression of the superconducting anomaly in magnetic fields from 1 to 13T in 1T increments.
Because the phonon specific heat is independent of magnetic field, the change in electronic specific heat with field is obtained by subtracting the zero field data from the data measured in a field (see Fig.~\ref{GSFFIG}(a)).
\begin{equation}
\Delta \gamma(H,T)=\gamma(H,T)-\gamma(0,T)
\label{DGEQN}
\end{equation}
Integrating over temperature gives the change in electronic entropy with field (see Fig.~\ref{GSFFIG}(b)).
\begin{equation}
\Delta S(H,T)=\int^T_0{\Delta \gamma(H,T)dT}
\label{ENTROPYEQN}
\end{equation}
The change in free energy is obtained by integrating the entropy over temperature (see Fig.~\ref{GSFFIG}(c)).
\begin{equation}
\Delta F(H,T)=-\int^T_0{\Delta S(H,T)dT}+\Delta F(H,0)
\label{FEQN}
\end{equation}
$\Delta F(H,0)$ is determined from the condition that the superconducting contribution to $\Delta F(H, T)$ tends to zero for $T\gg T_c$. 
At each temperature, the field dependence of the free energy is fitted to a theoretical expression derived from the model of Hao and Clem\cite{HAO} for an $s$-wave superconductor.
\begin{eqnarray}
\Delta F_{s,rev}(H,T)&=&\frac{a\phi_0}{32\pi^2\lambda^2}H\ln\left(\frac{e\beta H_{c2}}{H}\right)\\
\nonumber\\
&=&a_1H\ln(a_2H)
\label{HAOCLEMEQN}
\end{eqnarray}
The coefficients $a$ and $\beta$ are weakly field dependent and are given by $a\sim 0.77$ and $\beta\sim 1.44$ in the range $0.02<H/H_{c2}<0.3$.
A fit to the free energy at 2.5K is shown in the inset to Fig.~\ref{GSFFIG}(c).

\begin{figure}
\centering
\includegraphics[width=70mm,clip=true,trim=0 0 0 0]{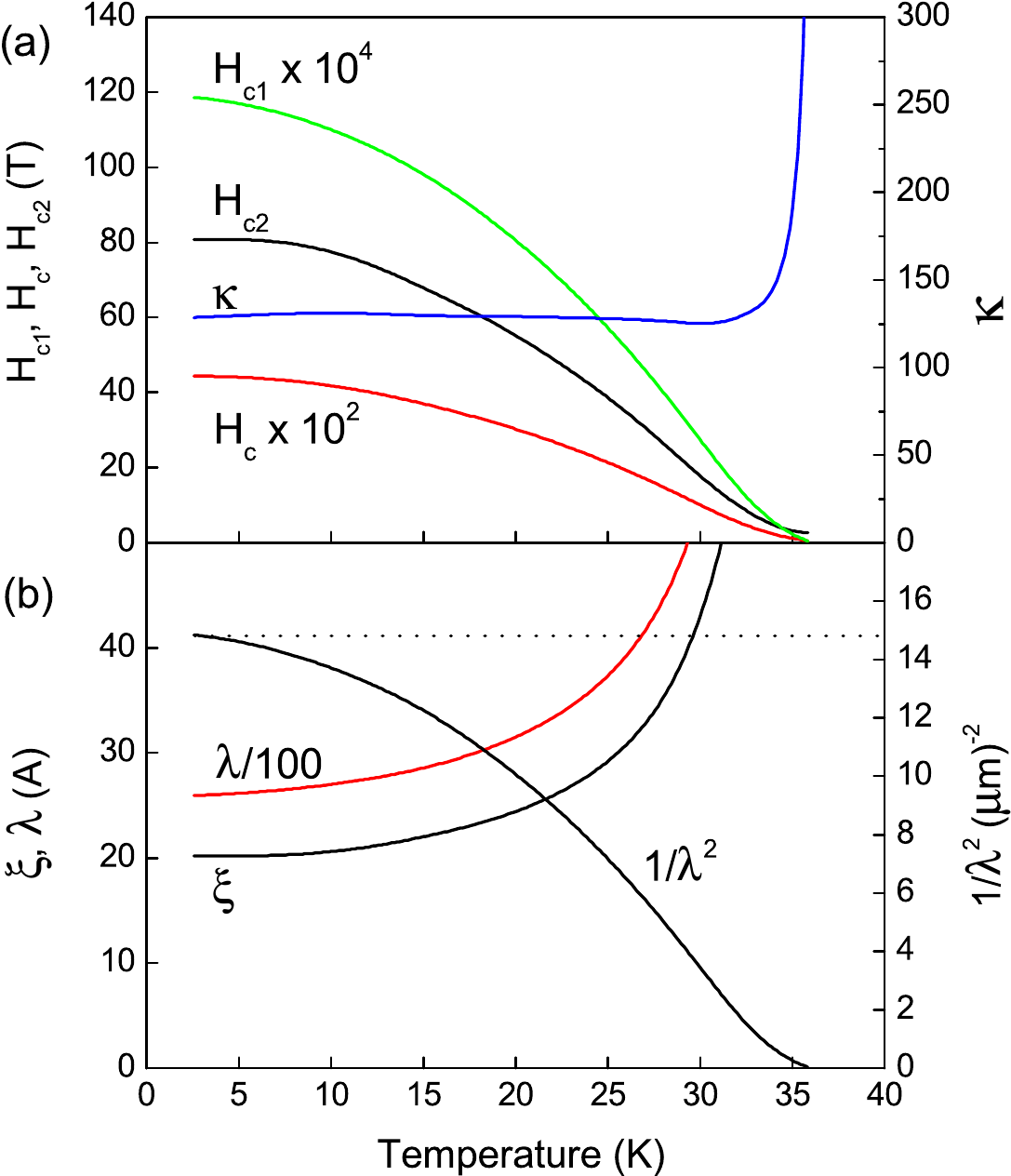}
\caption{Pollycrystalline averaged mixed state parameters of Ba$_{0.3}$K$_{0.3}$Fe$_2$As$_2$ derived from the field dependence of the free energy $\Delta F(H,T)$: (a) Critical fields $H_c$, $H_{c1}$, $H_{c2}$, and Ginsburg-Landau parameter $\kappa$; (b) London penetration depth $\lambda$, superfluid density $\propto 1/\lambda^2$ and superconducting coherence length $\xi$.}
\label{GLPARAMSFIG}
\end{figure}

The penetration depth, $\lambda$, and the upper critical field, H$_{c2}$, are determined directly from the fit parameters $a_1$ and $a_2$.
Then using the following Ginsburg-Landau relations we extract the:
critical field $H_c^2=(\phi_0/4\pi\lambda^2)H_{c2}$, Ginsburg-Landau parameter $\kappa=H_{c2}/H_c\sqrt{2}$, lower critical field $H_{c1}=(\phi_0 \ln{\kappa})/(4\pi\lambda^2)$, and superconducting coherence length $\xi=(\phi_0/2\pi H_{c2})^{1/2}$. The temperature dependencies of these quantities is plotted in Fig.~\ref{GLPARAMSFIG}. The values shown are polycrystalline averages. The large $\kappa\approx$ 130 indicates the strong type II nature of this material. Our $H_{c2}$ values are very similar to those obtained from radio frequency penetration depth measurements on a single crystal with $x$ = 0.45\cite{ALTARAWNEH}, and the value we obtain for $\lambda(0)\approx$ 260nm is similar to those measured by infrared spectroscopy\cite{LI} and tunnel diode resonator\cite{PROZOROV} techniques.

In summary, above 150K the electronic specific heats for the three samples are large and almost identical. For the $x$ = 0 sample there is a large and almost first order anomaly at the SDW transition at $T_0$ = 136K, whilst the corresponding anomaly for the $x$ = 0.1 sample at $T_0$ = 135K is smaller and broader and more closely resembles a second order phase transition. At low temperatures, $\gamma$ for these two samples is strongly reduced by the SDW gap by factors of 11 and 5 for $x$ = 0 and 0.1 respectively. 
In the $x$ = 0.3 sample the large superconducting anomaly and associated value of $S/T(T_c)$ shows that the underlying normal state $\gamma$ at low temperature is close to its high temperature value.  This suggests that as $x$ reduces, the growth of the SDW gap and consequent reduction in the density of states is responsible for the disappearance of superconductivity. The temperature dependence of  $\gamma$ ($x$ = 0.3) supports a nodeless superconducting gap function that is strongly anisotropic about the Fermi surface, possibly due to the presence of multiple gaps.

We gratefully acknowledge funding from the Engineering and Physical Sciences Research Council (U.K.) and the Swiss National Science Foundation pool MaNEP.

\end{document}